\documentstyle[aps,prl,preprint]{revtex}
\tightenlines
\draft
\begin{document}
\title {(Quasi)-Convexification of Barta's (Multi-Extrema) Bounding Theorem:
 $Inf_x\Big({{H\Phi(x)}\over {\Phi(x)}} \Big ) \leq E_{gr}
\leq Sup_x \Big({{H\Phi(x)}\over {\Phi(x)}}\Big) $}
\author{C. R. Handy  }
\address{
Department of Physics \& Center for Optimization Studies in the Applied
Sciences \\ Texas Southern University \\ Houston, Texas 77004 }
\date{Received \today}
\maketitle
\begin{abstract}
There has been renewed interest 
 in the exploitation of Barta's  {\it configuration space} theorem (BCST, (1937))
  which bounds  the  ground state
energy,
 $Inf_x\Big({{H\Phi(x)}\over {\Phi(x)}} \Big ) \leq E_{gr}
\leq Sup_x \Big({{H\Phi(x)}\over {\Phi(x)}}\Big)$, 
by using any $\Phi$ lying within the space of positive, 
bounded, and sufficiently smooth functions, ${\cal C}$. Mouchet's (2005) BCST analysis  
is based on gradient optimization (GO). However, it
 overlooks significant difficulties:
 (i) appearance of  multi-extrema;
(ii) inefficiency of GO  for stiff (singular perturbation/strong coupling) problems;
(iii) the nonexistence of a systematic procedure for arbitrarily improving the bounds
within ${\cal C}$. These deficiencies  can be corrected by transforming BCST
 into a moments' representation equivalent, and exploiting    a generalization of the
  Eigenvalue Moment Method (EMM), within the context of the well known Generalized Eigenvalue Problem (GEP),
 as developed here. EMM is an alternative
  eigenenergy  bounding, variational procedure, overlooked by Mouchet, which also exploits the positivity of 
the desired physical solution. Furthermore, it is  applicable to hermitian and non-hermitian systems with 
complex-number quantization parameters
(Handy and Bessis (1985), Handy et al (1988),
Handy (2001), and Handy, Msezane, and Yan (2002)).
Our analysis exploits various quasi-convexity/concavity theorems common to the GEP representation.
 We outline the general theory, and present some illustrative examples.
\end{abstract}

\vfil\break
\section{Introduction}
\subsection {Deficiencies of Barta's Configuration Space Formulation}
The recent work by Mouchet (2005) develops gradient optimization strategies for implementing Barta's (1937)
 eigenenergy bounding
procedure for the (bosonic) ground state energy, $E_{gr}$.
 In particular, for any
stricly positive trial function, $\Psi > 0$, within the set of positive,  twice differentiable, and 
bounded functions, ${\cal C}$, the infimum and supremum, as defined below,
generate  lower and 
upper bounds to the ground state energy:

\begin{equation}
Inf_x \Big({{H\Psi(x)}\over {\Psi(x)}} \Big ) \leq E_{gr}
\leq Sup_x \Big({{H\Psi(x)}\over {\Psi(x)}}\Big).
\end{equation}
 The restriction to the ground state is because of the well known theorem that in
configuration space, the ground state wavefunction  is  positive, $\Psi_{gr} > 0$.

Mouchet's analysis assumes that one is working with
 closed form expressions for the
trial wavefunction. Although this approach is quite flexible, it is an incomplete
resolution of an important, well known, deficiency of Barta's formalism (i.e. ``Barta's deficiency'').
 Specifically, given bounds generated from an initial trial function, how
does one systematically improve upon this, to arbitrary tightness of the
bounds, within the set  ${\cal C}$? This is particularly important since
the selection of suitable trial functions is not an intuitive process.
To begin to answer this, within Mouchet's formalism, one requires
infinite parameter representations for all of ${\cal C}$. This is
not possible in configuration space.

Besides this limitation, Barta's configuration space formulation also
presents significant multi-extrema  complications in sampling the
 ratio 
${{H\Psi(x)}\over \Psi(x)}$ over all $x$ values. Clearly, any alternate
approach that convexifies this problem (or its equivalent), thereby requiring
the determination of one global extremum, would be a tremendous improvement.
There are many important problems in physics where the appearance of multi-minima,
and strategies for circumventing these, is a major concern. One important
class of methods for doing this  simulated annealing (Kirkpatrick,  Gelatt,
and Vecchi (1983)). Related methods such as {\it iterative annealing} have impacted
more contemporary research such as those studying the  protein folding problem (Thirumalai and Hyeon (2005)).

An additional difficulty with  Mouchet's gradient approach is that it may not be the ideal strategy for
dealing with stiff systems (such as those associated with singular perturbation/strong coupling interactions)
 for which  very small integration steps
may slow the global search for the infimum and supremum. (By way of contrast,
the approach presented here does not require a gradient search. Instead a linear
programming based bisection approach proves highly effective.)

\subsection {A Remedy to Barta's Deficiencies: The Eigenvalue Moment Method}

In the 1980's these deficiencies were well known to D. Bessis and his group at Saclay,
particularly as a consequence of Barnsley's (1978) earlier studies on Barta's theorem. Their
objective was to  develop tight (converging) bounds for the notoriously
difficult quadratic Zeeman effect for hydrogenic atoms in superstrong magnetic fields (QZE).
In particular, Le Guillou and Zinn-Justin (1983) emphasized an intricate order dependent, hypervirial,
conformal analysis which gave numerical predicitions for the QZE ground state binding energy.
The QZE is an example of a singular perturbation/strong coupling problem (Bender and Orzag (1978)), for which different
methods (i.e. variational, numerical, analytical, etc.) can yield widely varying results.
Bessis was interested  in developing positivity  based  alternatives to Barta's theorem  
by which to asses the accuracy of 
of Le Guillou and Zinn-Justin's results.

Independent of these concerns, Handy (1984) discovered that certain well known theorems within the classic 
mathematical literature known, collectively, as the {\it Moment Problem} (Shohat and Tamarkin (1963)) could
be used to quantize physical systems through the generation of (geometrically) converging,
lower and upper bounds, to the ground state energy. The {\it Moment Problem} is 
concerned with the necessary and sufficient conditions the power moments of a function, $\mu_p = \int \ dx \ x^p \Psi(x)$,
must satisfy in order
to conclude that the underlying function is positive, $\Psi > 0$. More generally, Handy's interest in  moments' quantization formulations 
originated from earlier studies that suggested their effectiveness in  studying the
multiscale dynamics of certain
singular perturbation-strong 
coupling problems  (Handy (1981)). That is, a moments representation implicitly defined a multiscale hierarchy of sensitivity to smaller
and smaller scale structures. This was a precursor to wavelet theory (Grossmann and Morlet (1984) and Daubechies (1988), and was used,
more recently, to incorporate continuous wavelet transform theory into quantum mechanics (Handy and Murenzi  (1997,1998,1999)).

Handy's eigenenergy bounding procedure exploited the confluence of several, hitherto, separate results. These were:
(i) the (multidimensional) Schrodinger equation with rational fraction potential could be readily
transformed into a {\it moment equation} recursion relation for the power moments of the wavefunction, $\mu_p = \int \ dx \ x^p \Psi(x)$,
involving the energy as a parameter, $E$; (ii) the (multidimensional) bosonic ground state wavefunction
must be positive, $\Psi_{gr} > 0$; (iii) the {\it moment problem}  positivity theorems define an infinite hierarchy of
constraints on the power moments, and in turn, on $E$. The particular positivity theorem
used by Handy (1984) was the well known nesting constraints for the (diagonal and off-diagonal) Pade approximants (generated from the $\mu_p$'s)
 of the associated
Stieltjes integral for $\Psi_{gr}$ (Baker (1975)). 

In their first collaboration, Handy and Bessis (1985) recognized that the Stieltjes-Pade theorems would not be extendable to
multidimensions. They proposed to exploit the {\it moment problem} positivity theorems based on the nonlinear, Hankel-Hadamard (HH)
determinantal inequalities. These could be extended, in principle, to multidimensions; however, this proved too costly. In a 
subsequent breakthrough (Handy et al (1988)), they realized that the HH nonlinear formulation could be transformed into an equivalent linearized version,
which was then amenable to linear programming analysis (Chvatal (1983)). Handy devised an efficient (bi-section) method 
referred to as the {\it Cutting Algorithm}, which led to the generation of tight, converging, bounds to the QZE problem, confirming
the results of Le Guillou and Zinn-Justin (Handy, Bessis, Sigismondi, and Morley (1988)). The entire procedure is referred to as the
Eigenvalue Moment Method (EMM).

Thus, historically, the EMM approach was specifically invented to bypass all of the unattractive features of  Barta's configuration
space  theorem. Mouchet overlooks this in his review of variational (bounding) methods, despite the fact that 
the EMM procedure is an {\it affine map invariant, variational procedure} (Handy and Murenzi (1998)). That is, the EMM bounds automatically 
sample over all affine map transformations (translations, scalings, rotations, etc.) of the trial  functions (i.e. polynomials). 
Affine maps are at the heart of fractals (Barnsley (1988))
and  wavelet transform theory (Grossman and Morlet (1984), Daubechies (1988)). Fractals and wavelets   represent important representations 
for dealing with systems with significant multiscale structures. For this same reason, the EMM bounds are very good  for dealing with (stiff) singular
perturbation type systems with signficant multiscale dynamics.

\subsection { Moment Problem Reformulation of Barta's Theorem: The Generalized Eigenvalue Problem}

Despite the successes of the EMM procedure, Mouchet's work has inspired the author to develop a {\it Moment Problem} counterpart to Barta's
configuration space theorem. Several new results ensue from this. The first is that the new formulation does not require the introduction
of a {\it moment equation}. Instead, we are able to generalize the underlying EMM philosophy in a manner  more in keeping with operator (matrix)
theory. Whereas the moments in the EMM formulation are constrained by the moment equation relations, in the new formulation, all of the 
moments are unconstrained relative to one another. The energy parameter does not explicitly appear. Instead, one studies the extremal
eigenvalues of the corresponding {\it Generalized Eigenvalue Problem} (GEP) (Boyd and Vandenberghe (2004)) defined according to
${\bf H} {\overrightarrow V} = \lambda {\bf U} {\overrightarrow V}$,
where ${\bf H}$ represents the moment representation operator (matrix) for the Hamiltonian, and ${\bf U}$ represents a positive definite 
matrix operator (the Hankel moment matrix).
 We are not interested in the generalized eigenvectors , ${\overrightarrow V}$, but make reference to them for clarity.
Instead, it is the extremal  eigenvalues, $\lambda_{min} \leq \lambda \leq \lambda_{max}$, that are of interest, since these will directly relate to the infimum and
supremum expressions in Barta's configuration space theorem. 

We will investigate two different versions of the above GEP system. In the first case, we will assume that the moments, ${\mu_p}$, used in defining the ${\bf (H,U)}$
operator pair correspond to a positive trial function (i.e. either we have a closed form for the function, $\Psi$, and its power moments, $\mu_p$;
 or we only know the $\mu_p$'s, but have no closed form expression for $\Psi$).
 We can then generate monotonically converging sequences to both the infimum and supremum:
  $Inf_x \Big({{H\Psi}\over \Psi}\Big) < \ldots < \lambda_{min;n} < \ldots < \lambda_{min;1}$ and
$ \lambda_{max;1} < \ldots < \lambda_{max;n} < \ldots < Sup_x\Big({{H\Psi}\over \Psi}\Big) $. 
In principle, this already
defines one clear advantage, since any multi-extrema features of the ratio ${{H\Psi}\over \Psi}$ are circumvented by     
our ability to generate                                      
monotonically converging sequences. 

The $({\bf H},{\bf U})$  matrix operator pair, although of infinite dimension, can be studied in terms of their finite dimensional, upper left hand,
submatrices. For each of these submatrices, we can define a finite dimensional, convex moment variable space, ${\cal U}_n$. In our second
study of the GEP formulation, we will determine the optimal values, over  ${\cal U}_n$, for each of the extremal GEP eigenvalues 
 (i.e. $Sup_{\mu \in {\cal U}_n}(\lambda_{min;n}(\mu))$ and $Inf_{\mu \in {\cal U}_n}(\lambda_{max;n}(\mu)$)). These will generate converging
bounds to the ground state energy: $Inf_{\mu \in {\cal U}_n}(\lambda_{max;n}(\mu)) < E_{gr} < Sup_{\mu \in {\cal U}_n}(\lambda_{min;n}(\mu))$.

\subsection {Technical Preliminaries}

In order to clarify the notation used in connection with the new contributions of this work, we
develop them in the context of a short technical overview of the Eigenvalue Moment Method (EMM). The EMM will also play
an important role in proving some of the theorems introduced in this work; therefore, its review, here, will facilitate 
the overall understanding of our new results.

\subsubsection {The Moment Equation}

All of our results are predicated on being able to transform the Schrodinger equation
into a {\it Moment Equation} representation. This is always possible for multidimensional
systems with rational fraction potentials. For problems not of this type, it still may
be possible to identify coordinate systems in which the transformed Schrodinger equation
involves function coefficients that are of rational fraction form. This was the case for
the quadratic Zeeman effect, as previously cited, for which parabolic coordinates led to
the identification of a moment equation. Despite these limitations, these restrictions
still define a large and important class of physics problems. 

Because our intent is to develop the underlying theory, we have chosen
to limit all discussions, and examples, to one dimensional systems, for simplicity.

 For 
one dimensional systems, the moment equation corresponds to
a recursive, linear, homogeneous,  finite difference equation of order $1+m_s$, wherein all
of the moments are linearly dependent on the first $1+m_s$ moments $\{\mu_0,\ldots,\mu_{m_s}\}$.
The latter will be referred to as 
the {\it initialization} moments, or 
 the {\it missing moments}.
 Once any suitable normalization prescription is adopted, 
the moments can be written in terms of the unconstrained initialization moments. The form of the
normalization prescription will vary, depending on the nature of the problem (i.e. Stieltjes, Hamburger, etc.).
In the Hamburger case, the even  moments must be positive, the odd moments can have arbitrary signature.
We want to choose a normalization prescription that automatically bounds the initialization  moments. For the Hamburger
case, one must have $m_s = even$, and we can take $\mu_0 + \mu_{m_s} = 1$.  It then follows 
that $|\mu_\ell| \leq \mu_0 + \mu_{m_s} = 1$, for $0 \leq \ell \leq m_s$. Since $\mu_0 = 1 - \mu_{m_s}$,
all of the remaining initialization moments are unconstrained. We can now write the moment equation
as

\begin{equation}
\mu_p = {\hat M}_E(p,0) + \sum_{\ell = 1}^{m_s} {\hat M}_E(p,\ell)
 \mu_\ell, \ p \geq 0,
\end{equation}
where the ${\hat M}_E(p,\ell)$ are known coefficients, nonlinearly dependent on the energy, $E$. We note
that the (unconstrained) initialization moments must lie within the $m_s$ dimensional cube:
$(\mu_1,\ldots,\mu_{m_s}) \in (-1,1)^{m_s}$.

\subsubsection {Stieltjes-Pade Positivity Quantization}

For parity invariant systems,  the change of variables $x = \sqrt{y}$,
 $y \geq 0$, transforms the Hamburger moment problem (involving functions on the entire real axis, $\Re$, $\mu_p = \int_{-\infty}^{+\infty}dx \ x^p \Psi(x)$) into a Stieltjes moment problem (functions restricted to the nonnegative real axis, $u_p = \int_0^\infty \ dy \ y^p\Phi(y)$). For some of these systems, such as the 
harmonic oscillator problem (i.e. $- \partial_x^2\Psi + x^2\Psi(x) = E\Psi(x)$ ), the moment equation's order becomes unity, or $m_s = 0$. In such cases, the even order Hamburger moments of the wavefunction become
Stieltjes moments of the modified wavefunction $\Phi(y) \equiv {{\Psi(\sqrt{y})}\over {\sqrt{y}}}$  : $\mu_{2\rho} = u_\rho  = \int_0^\infty dy \ y^\rho \Phi(y)$. Because $m_s = 0$, the Stieltjes moments for the ground state become (known) nonlinear 
functions of $E$, the energy variable ( i.e. $u_{\rho+1} = E u_\rho + 2\rho(2\rho-1)u_{\rho-1}$, $u_0 \equiv 1$, $u_1 = E$, $u_2 = 2 + E^2$, etc.).

Since the Stieltjes moments for the $\Phi$-ground state are known functions of $E$, 
 these then also 
determine the Pade approximants (Baker (1975)), $[M|N]_{(E;s)}$, for the associated Stieltjes
integral, 
 $I(s) = \int_0^\infty dy \ {{\Phi(y)}\over{1+sy}}$.

It is a well known theorem
that if a Stieltjes measure is positive, then the $[M|M]$ and $[M-1|M]$
Pade approximants must satisfy a nested structure:

\begin{equation}
 [M-1|M]_{(E_g;s)} \leq [M|M+1]_{(E_g;s)} \leq I(s) \leq
 [M+1|M+1]_{(E_g;s)} \leq [M|M]_{(E_g;s)} 
\end{equation}

Handy (1984) discovered that
one could use this nested behavior to quantize the ground state energy, 
through converging lower and upper bounds. Thus, for arbitrary $E$, one
generates the first $Q$ Stieltjes moments (and all the Pade approximants 
that can be generated from them), and determines the energy interval,
 $(E_Q^{L}, E_Q^{U})$, of feasible energy values that lead to Pade
approximants satisfying the 
 above nested  structure. The endpoints
of the  feasibility energy interval become the numerically generated
lower and upper bounds to the ground state energy: $E_Q^{L}\leq E_{gr}
 \leq E_Q^{U}$. The entire process
is repeated at the next higher order ($Q \rightarrow Q+1$), resulting in a reduction of the
feasibility energy interval. In this manner, geometrically converging,
 lower and upper 
bounds are obtained.

\subsubsection {Hankel-Hadamard Determinant Positivity Quantization}

Since Pade approximants could not  be easily extended to multidimensions, 
 an alternate equivalent to the above Moment Problem quantization
procedure was required. The standard Moment Problem positivity 
constraints, for a nonnegative function,
 $f(x) \geq 0$
(excluding distribution type
expressions with zero measure support), are generally expressed in
terms of the Hankel-Hadamard (HH) determinantal, inequality constraints, given
in Eq.(6). These are derived through the quadratic form integral expression

\begin{equation}
\int_{-\infty}^{+\infty} dx \ \Big(\sum_{n = 0}^N C_n x^n\Big)^2 \ f(x) > 0,
\end{equation}
or
\begin{equation}
\langle {\overrightarrow C}| \pmatrix { \mu_0, \mu_1, \ldots, \mu_N \cr
\mu_1,\mu_2, \ldots, \mu_{N+1} \cr
\cdots \cr
\mu_{N}, \mu_{N+1}, \ldots, \mu_{2N} \cr } | {\overrightarrow C} \rangle > 0, 
\ \forall {\overrightarrow C} \neq {\overrightarrow 0}. \end{equation}
 The real and symmetric Hankel moment matrix (defined in Eq.(5)) is 
therefore positive definite, with positive eigenvalues. Thus all of its
subdeterminants of the following type, must be positive:

\begin{equation}
\Delta_{m,n}\Big( \mu \Big ) > 0, m_{ (even)} \geq 0, n \geq 0, 
\end{equation}
where 

\begin{equation}
\Delta_{m,n}\Big( \mu \Big ) \equiv Det  \pmatrix { \mu_m, \mu_{m+1}, \ldots, 
\mu_{m+n} \cr
\mu_{m+1}, \mu_{m+2}, \ldots, \mu_{m+n+1} \cr
\cdots \cr
\mu_{m+n}, \mu_{m+n+1}, \ldots, \mu_{m+2n} \cr}.
\end{equation}
For the Hamburger Moment Problem (i.e. the moment constraints leading to 
a positive function on $\Re$), it is sufficient to 
to require that $\Delta_{m,n}\Big( \mu \Big ) > 0$ for $m = 0, n \geq 0$ (Baker (1975)).  

It is clear that the HH inequalities are necessary conditions for any positive
function. That they are sufficient for establishing the positivity
of the underlying function can be motivated as follows.
One can approximate the gaussian, dirac 
distribution, in terms of the quadratic form expansion. That is 
${1\over{\beta\sqrt{\pi}}}e^{-{{(x-\tau)^2}\over \beta}} =
{1\over{\beta\sqrt{\pi}}} \Big(e^{-{{(x-\tau)^2}\over{2 \beta}}}\Big)^2 \approx
{1\over{\beta\sqrt{\pi}}}\Big( \sum_{j=0}^J {1\over{j!}}\Big({-{{(x-\tau)^2}\over{2 \beta}}}\Big)^j\Big)^2$. Thus, for sufficiently large $J$ and small $\beta$ values, the HH constraints are essentially sampling the local behavior of 
a (bounded, asymptotically decaying) function, and requiring that it be positive. 

The HH determinants can be extended to multidimensions, as developed in
the work by Devinatz (1957).

 The first
use of the HH, Moment Problem (MP), positivity theorems to quantize the bosonic
ground state energy was published by Handy and Bessis (1985). We briefly
outline the essentials of this work.
Through the Moment Equation's structure, Eq.(2),  the moments are explicitly dependent on the energy variable, $E$,
and the (unconstrained) {\it initialization moments}, $\{\mu_1,\ldots, \mu_{m_s}\}$. So too
are the HH determinants, $\Delta_{0,n}(\mu) = \Delta_{0,n}(\mu_1,\ldots, \mu_{m_s};E )$, for $n < \infty$.
 Given an 
arbitrary $E$ value, and (even number) moment expansion order, $Q < \infty$ (thus all the moments $\mu_{p \leq Q}$ are generated)
 one determines if there
exists an $m_s$-dimensional,{\it initialization moment} solution set, ${\cal U}_{Q;E} \subset (-1,1)^{m_s}$, satisfying all the corresponding HH inequalities ($ \Delta_{0,0}(\mu) > 0, \Delta_{0,1}(\mu) > 0, \ldots, \Delta_{0,{Q\over 2}}(\mu) > 0$).
 If this solution set exists (${\cal U}_{Q;E} \neq \oslash$), then the associated $E$ value
is a possible physical ground state value, to order $Q$. If not (${\cal U}_{Q;E} = \oslash$), then the chosen $E$ value  is not a possible physical value for the ground state energy.
In this manner, a feasibility energy interval is determined, as before, in
the Pade case. It can be shown that ${\cal U}_{Q;E}$ must be a convex set,
if it exists ($E \in (E_Q^{L}, E_Q^{U}) \iff {\cal U}_{Q;E} \neq \oslash$).

\subsubsection {Variational-Linear Programming, Moment Problem Quantization}

Although the previous HH-MP procedure created greater flexibility in 
extending the underlying positivity quantization philosophy to more systems, its structure
(i.e. the nonlinear dependence on the moments) made it too difficult for 
multidimensional systems. 

Any bounded, convex, set (with nonlinear boundaries), such as the
${\cal U}_{Q;E}$'s can be represented as the intersection of infinitely
many bounded polytopes (convex sets with hyperplanes as boundaries). 
One realizes that Eq.(5) defines this equivalent, alternative linear representation for the HH inequalities. That is, instead of working with a finite number
of (large dimensioned) nonlinear inequality relations (i.e. the HH constraints),one can work with an equivalent set of  infinitely many, linear, constraints.
This provided the  theoretical breakthrough in facilitating the
implementation of Moment Problem positivity quantization strategy.

In order to capitalize on this linearized equivalent, one must devize an 
optimization strategy to essentially determine the optimal ${\overrightarrow C}$'s, or {\it cutting vectors} (refer to Eq.(5)). This required a clever combination of the 
moment equation formalism with linear programming theory (Chvatal (1983)).
 Thus, through the ensuing {\it Cutting Algorithm}, devized by Handy, given an 
arbitary $E$ value, one rapidly ``cuts-up'' 
  the starting (normalization) polytope ( the  hypercube, $(-1,1)^{m_s}$) into either the null set
(thereby establishing that ${\cal U}_{Q;E} = \oslash$ and  $E$ is unphysical), or into a polytope, ${\cal P} \supset {\cal U}_{Q;E}$, containing an {\it initialization point}, $\overrightarrow {\tilde \mu} = ({\tilde\mu}_1, \ldots, {\tilde \mu}_{m_s}) $ for which all the associated Hankel matrices are 
positive (concluding that $\overrightarrow {\tilde \mu} \in {\cal U}_{Q;E}$, hence
  ${\cal U}_{Q;E} \neq \oslash$, and that particular $E$ is
a possible value for the ground state energy).

The above, entire, procedure is  
the Eigenvalue Moment Method (EMM), which was used to solve the previously cited quadratic Zeeman problem.
\vfil\break
\section {Moment Problem Reformulation of Barta's  Theorem \\ as a Generalized 
Eigenvalue Problem}

In this work we directly
 transform Barta's configuration space theorem into
a Moment Problem representation. By so doing, we
obtain a theoretically complete formalism that addresses and
solves Barta's configuration space deficiencies. Our analysis is limited to Hamiltonian systems
which can be represented in terms of a moment representation.
 As previously noted,
this corresponds to a large, and important, class of physically interesting
systems.

In contrast to the EMM approach which restricts itself to the set of moments
satisfying the physical moment equation, the new formalism makes no such
restriction. 
 Previously, ${\cal U}_Q$ referred to a subset  within the domain of
initialization moments (i.e. the subset of initialization moment values whose
Moment Equation generated moments, up to moment order $Q$, satisfy the HH 
positivity constraints):

\smallskip\noindent {\bf{ Definition (Sec. I)}}
 
\smallskip\noindent  ${\cal U}_Q = \{ (\mu_1,\ldots,\mu_{m_s}) |
\mu_p = {\hat M}_E(p,0) + \sum_{\ell = 1}^{m_s} {\hat M}_E(p,\ell)
 \mu_\ell,\ 0 \leq p \leq Q,\ and\ \Delta_{0,n}(\mu) > 0,\ 0 \leq n \leq {Q\over 2},  \} \subset (-1,1)^{m_s}$

Note that the elements of ${\cal U}_Q$ implicitly must satisfy some, physically motivated, normalization prescription. 

The new definition of this same notation will be

\smallskip\noindent {\bf{ Definition (Modified)}} 

\smallskip\noindent  ${\cal U}_Q = \{ (\mu_0,\ldots,\mu_Q)|                                                                                                 
 where\ \Delta_{0,n}(\mu) > 0,\ 0 \leq n \leq {Q\over 2}\}$.

That is, ${\cal U}_Q$
 refers to the domain of
Hamburger moments, up to moment order, $Q$, satisfying all the 
corresponding HH positivity 
constraints (and not constrained to satisfy any moment equation). In either case, $Q$ is implicitly an even number.

It is also implicitly assumed that the elements of
${\cal U}_Q$ must satisfy some physically motivated normalization prescription.

Within each of the finite dimensional subsets, ${\cal U}_Q$,
Barta's relations manifest themselves in terms of a {\bf Generalized Eigenvalue 
Problem} (GEP) (Watkins (2002))
\begin{equation}
 {\bf H}|{\overrightarrow V}\rangle = \lambda {\bf U}|{\overrightarrow V}\rangle,
\end{equation}
where {\bf H} and {\bf U} correspond to finite, real and symmetric  matrices, 
to be defined in Sec. III.
The matrix elements will  be linear in the moments.
 Because of the 
restriction to ${\cal U}_Q$ 
 the {\bf U}-Hankel matrix is positive 
definite.
 The matrices ({\bf H},{\bf U}) are designated as a {\it symmetric} pair. As will be seen by the
explicit example discussed below, whereas all of the moment variables in
${\cal U}_Q$ contribute to the structure of ${\bf H}$, a reduced number of
these contribute to ${\bf U}$; however, this reduced number still guarantee
the positive definiteness of ${\bf U}$.

Through a Cholesky decomposition (Watkins (2002)), {\bf U} = {\bf R}$^t${\bf R}, the GEP
is transformed into a standard, symmetric matrix, eigenvalue problem,

\begin{equation}
{\bf R}^{-t} {\bf H}{\bf R}^{-1} |{\overrightarrow W}\rangle = \lambda |{\overrightarrow W}\rangle
.
\end{equation}

The extremal eigenvalues $\lambda_{min} \leq \lambda \leq \lambda_{max}$,
satisfying the above, or alternatively $Det({\bf H} - \lambda {\bf U}) = 0$, 
are defined by:

\begin{equation}
\lambda_{min}(\mu) = Inf_{\overrightarrow C}\ {{\langle {\overrightarrow C}| {\bf H}(\mu) | {\overrightarrow C}\rangle}\over{{\langle {\overrightarrow C}| {\bf U}(\mu)
| {\overrightarrow C}\rangle}}}, \ \{\mu\} \in {\cal U}_Q.
\end{equation}
and
\begin{equation}
\lambda_{max}(\mu) = Sup_{\overrightarrow C}\ {{\langle {\overrightarrow C}| {\bf H}
(\mu) | {\overrightarrow C}\rangle}\over{{\langle {\overrightarrow C}| {\bf U}
(\mu)
| {\overrightarrow C}\rangle}}}, \ \{\mu\} \in {\cal U}_Q.
\end{equation}
Of course, we also have
 $\lambda_{min}(\mu) =  -Sup_{\overrightarrow C}\
 {{\langle {\overrightarrow C}| {\bf -H}
(\mu) | {\overrightarrow C}\rangle}\over{{\langle {\overrightarrow C}|
 {\bf U}(\mu)
| {\overrightarrow C}\rangle}}}$.

The above ratios can also be written in terms of 
${  { {\langle {\overrightarrow C}| {\bf H}(\mu) | {\overrightarrow C}\rangle  }\over {\langle {\overrightarrow C}|{\overrightarrow C}\rangle }}
\over { {\langle {\overrightarrow C}| {\bf U}(\mu)| {\overrightarrow C}\rangle}   \over {\langle {\overrightarrow C}|{\overrightarrow C} \rangle } }   }$.
Let $\lambda_{\bf H,\bf U}^{min,max}(\mu)$ denote
 the extremal eigenvalues of the ${\bf H}$ and ${\bf U}$ matrices, respectively. Since
$\lambda_{\bf U}^{min}(\mu) > 0$, if we also assume that  $\lambda_{\bf H}^{min}(\mu) > 0$ (for simplicity), then
  ${{\lambda_{\bf H}^{min}(\mu)}\over {\lambda_{\bf U}^{max}(\mu)}} \leq \lambda_{min}(\mu) \leq \lambda_{max}(\mu)
 \leq {{\lambda_{\bf H}^{max}(\mu)}\over
{\lambda_{\bf U}^{min}(\mu)}}$.  However, if
 $\lambda^{min}_{\bf H}(\mu) < 0$, then ${{\lambda_{\bf H}^{min}(\mu)}\over {\lambda_{\bf U}^{min}(\mu)}} \leq \lambda_{min}(\mu) $. These expressions
will prove important later on.

Physicists refer to any function which, locally, lies
 below the tangent plane  as a
{\it convex function}.  Mathematicians define  a function,
 $f(x)$, as  convex  if the set $\{(x,y)| y \geq f(x)\}$ is convex. Thus, what a physicist would regard as a concaved function, is referred to as 
a convex function by mathematicians. Throughout this work we will use the
physicist's definition, except in a few cases where we cite the actual theorems (placing quotation marks
around them), as they appear in the literature.

{\bf Definition}

\bigskip\centerline{$ \pmatrix{Concaved \cr Convex} Function_{[physicists]} \equiv 
 \pmatrix{``Convex" \cr ``Concaved"} Function_{[mathematicians]}$}

It is a well known theorem that the smallest eigenvalue of a symmetric matrix is a convex function
 with regards to the
matrix elements as variables. This led to various alternative 
algorithmic strategies (i.e. gradient methods) for implementing EMM (Handy, Giraud, and Bessis (1991), and Handy, Maweu, and Atterberry (1996)).

 The smallest eigenvalue of the generalized
eigenvalue problem is not convex as such, but it shares the good fortune
that it does not have
any multi-maxima or saddle points. That is, there can be regions of 
relative flatness. Thus, although there can only be one
 global maximum value,
the points in ${\cal U}_Q$ corresponding to the global maximum may not be 
unique. These properties are what the mathematicians refer to as {\it \underbar{quasi}-``concaved"}.

Despite this, 
in the infinite limit (all of moment space)  the 
physical problem strongly suggests that the point, in the moment variables space, 
corresponding to a global maximum for $\lambda_{min}(\mu)$,  is unique. This is because there can only 
be one physical ground state (the ground state energy can only be associated
with one point in ${\cal U}_\infty$). Since
we are only interested in obtaining tight bounds for $E_{gr}$ these 
issues do not affect this objective.
    
	We now revert to the mathematics nomenclature which is opposite to the physicist's
intuitive interpretation, as previously noted.

\noindent {\bf Definition}:
A function, $f({\overrightarrow \mu}): {\cal U} \rightarrow \Re$,
is quasi-``concaved" if 

\begin{equation}
 f(s {{\overrightarrow \mu}_1} + (1-s)
{{\overrightarrow \mu}_2}) \geq min\{ f({{\overrightarrow \mu}_1}),
f({{\overrightarrow \mu}_2})\}, \ 0 \leq s \leq 1, {{\overrightarrow \mu}_{1,2}} \in {\cal U}.
\end{equation}

It therefore follows that for a quasi-``concaved" function there can be flat regions where the function stays constant. However,  along the one dimensional 
path defined by $0 < s < 1$, there can be no local minimum. If the function
is strictly ``concaved", 
 then any local differential search will always
yield a path which leads to the global maximum. However, for quasi-``concaved" functions, within regions of flatness, more effort may be required to find 
a path that leads to a global maximum.

We have the important theorem (mathematical nomenclature)

\noindent {\bf Theorem \# 1}:
$\lambda_{max}(\mu)$   is a {\bf quasi-``convex"} function
of the matrix elements ({\bf H},{\bf U}),
 which are linear in the moments (Siddharth (2005), Boyd and Vandenberghe (2004))
.

Similarly,
$\lambda_{min}(\mu)$ (being the negative of a quasi-``convex"
function, refer to discussion following Eq.(11)) is a  {\bf quasi-``concaved"} function.

From the definition of the extremal eigenvalues
one has that $\lambda_{min}(\mu) \leq \lambda_{max}(\mu)$. However, there
will be moment elements in ${\cal U}_Q$ for which the extremal 
eigenvalues coincide. This will be the case for those satisfying the 
moment equation, up to moment order $Q$ (i.e. those moments that also satisfy the EMM moment equation).

\bigskip\noindent {\bf Definition}:
Denote by $ {\overrightarrow \mu}_E = \{\mu_0,\mu_1,\ldots,\mu_{Q}\} \in {\cal U}_Q$, an element of ${\cal U}_Q$ that also
 satisfies the moment equation (Eq.(2)), for the given $E$ value. Such a point, by definition,
  automatically
 satisfies the EMM positivity
constraints. This is only possible if  $E \in (E^L_{Q},E^U_{Q})$. That is, it must lie within the EMM eigenenergy bounds. Now  use it
to generate the $({\bf H},{\bf U})$ symmetric pair matrices, as defined
in the next section. It will be shown in the following section that

\noindent{\bf Theorem \#2}
\begin{equation}
\lambda_{min}(\mu_E) = E =  \lambda_{max}(\mu_E).
\end{equation}
We shall denote by ${\cal U}_{Q;EMM} \subset {\cal U}_Q$, the subset of
points satisfying the EMM conditions (i.e. satisfies the moment equation up
to order $Q$, and the positivity conditions, for an $E$ value that
must lie within the EMM bounds).

Let us explicitly distinguish the extremal GEP eigenvalues for each ${\cal U}_n$ of dimension $n+1$  by the notation:  $\lambda_{max/min;n}(\mu)$. 
Also, let ${\cal C}$ denote the set of functions that are positive, bounded (exponentially decreasing), and continuously differentiable up to
the second order. In the next section we shall show that if $\Psi \in {\cal C}$, 
and $\mu_p = \int dx \ x^p \Psi(x), p < \infty$
are its moments, then 

\noindent {\bf Theorem \# 3}:
\begin{equation} 
   Inf_x\Big( {{H\Psi(x)}\over {\Psi(x)}}\Big) \leq \lambda_{min;n+1}(\mu) \leq \lambda_{min;n}(\mu),
\end{equation}
\begin{equation}
\lambda_{max;n}(\mu) \leq \lambda_{max;n+1}(\mu) \leq Sup_x\Big( {{H\Psi(x)}\over {\Psi(x)}}\Big)  ,
\end{equation} 
and
\begin{equation}
\lim_{n\rightarrow \infty} \pmatrix { \lambda_{min;n}(\mu) \cr \lambda_{max;n}(\mu) \cr} = \pmatrix{ Inf_x\Big( {{H\Psi(x)}\over {\Psi(x)}}\Big) \cr Sup_x\Big( {{H\Psi(x)}\over {\Psi(x)}}\Big) },
\end{equation}
for each $\Psi \in {\cal C}$.

This is an interesting result, particulary when combined with sequence acceleration methods, since it allows one to determine  Barta's lower and upper bounds
for a function whose moments are known, even when the function is not given
in closed form. Also, the monotonic nature of the results may prove useful in circumventing potential multi-extrema features of the
${{H\Psi}\over \Psi}$ ratio, when evaluated in terms of Barta's configuration space formulation.

Let us now define the Sup and Inf of the extremal eigenvalues over their finite
dimensional convex domain, ${\cal U}_n$:

\begin{equation}
\pmatrix{\lambda_{min;n}^{Sup}\cr 
\lambda_{max;n}^{Inf}\cr } \equiv \pmatrix{ Sup_{\mu \in {\cal U}_n} \Big(\lambda_{min;n}(\mu) \Big) \cr
 Inf_{\mu \in {\cal U}_n}\Big( \lambda_{max;n}(\mu)\Big) \cr}.
\end{equation}

From Eq.(13), since ${\cal U}_{n;EMM} \subset {\cal U}_n$, we must have that 
$ \lambda_{min;n}^{Sup} \geq E^U_n$, the EMM upper bound. Similarly,
$ \lambda_{max;n}^{Inf} \leq E^L_n$, the EMM lower bound:

{\bf Theorem \# 4}

\begin{equation}
 \lambda_{max;n}^{Inf} \leq E^{EMM-lower bound}_n \leq E_{gr} \leq  E^{EMM-upper bound}_n \leq \lambda_{min;n}^{Sup},
\end{equation}
and, in the infinite limit,
 $n \rightarrow \infty$:

\begin{equation}
\lambda_{max;n}^{Inf} \leq \lambda_{max;n+1}^{Inf} \leq E_{gr} \leq 
\lambda_{min;n+1}^{Sup} \leq \lambda_{min;n}^{Sup}.
\end{equation}

\noindent{\bf Important Assumption/Condition} Although ${\cal U}_n$ will be a bounded convex set through the normalization conditions used,
it is also important that its boundary ($\delta {\cal U}$) not include points at which the positive definetness is lost.
 That is,
if ${\overrightarrow {\mu_b}} \in \delta {\cal U}$, we do not want $\Delta_{0,j \leq n}(\overrightarrow {\mu_b}) = 0$, or
$\lambda_{\bf U}^{min}({\overrightarrow {\mu_b}}) = 0$, using the notation in the discussion following Eq.(11). If this
is were to happen, then the $\lambda_{min;n}^{Sup}$, as well as $\lambda_{max;n}^{Inf}$, could become singular (i.e. $+\infty,-\infty$,
respectively). In the present application, we can insure the above by simply imposing additional moment inequality constraints
associated with any (rough) upper bound to the ground state energy. This will be clarified in the last section, where we implement
the numerical analysis on a specific problem.
\vfil\break
\section {Proof of Theorems}

In this section we develop the basic relations and prove the
various theorems previously quoted, with the exception of
Theorem \# 1 which is a standard result in optimization theory,
particularly in the context of mathematical economics. We will be 
limiting our discussion to the one dimensional case, for simplicity.

Two crucial elements are required for deriving  Barta's theorem in Eq.(1).
The first is that in the configuration space representation, the
bosonic ground state wavefunction, $\Psi_{gr}(x)$, must be of uniform
signature, and thus can be taken to be positive, $\Psi_{gr}(x) > 0$.
Accordingly, given any trial function, $\Psi$, of arbitrary signature, and
with a bounded and continuous second order 
derivative, one obtains the zero identity for the scalar product:
$\langle \Psi_{gr}| (H-E_{gr})|\Psi\rangle = 0$, where $E_{gr}$ is the ground state energy, and $H$ is the Schrodinger equation hamiltonian.
Therefore,
$H-E_{gr}$, when applied to 
$\Psi$, must have a zero at some location

\begin{equation}
\Big(H-E_{gr}\Big) \Psi(x_0) = 0.
\end{equation}

The second assumption is that if the trial function is stricly positive, $\Psi > 0$, then the range of the  function
 $R(x) = {{H\Psi(x)}\over{\Psi(x)}}$ must define a bounded subset of $\Re$
that contains $E_{gr}$: $E_{gr} \in \{ R(x) | \forall x \in \Re \}$. This
leads to Eq.(1) or ${\it Inf}_x R(x) \leq E_{gr} \leq {\it Sup}_x R(x)$.

Positivity is an important cornerstone of Barta's theorem.

Let ${\cal S}$ and ${\cal I}$ denote the supremum
and infimum, respectively, for an arbitrary trial function,
$\Psi$, lying within the set of functions,  $ {\cal C}$, which are
 positive, bounded (exponentially decaying), and have continuous, finite, second derivatives:

\begin{equation}
{\cal I} \equiv Inf \Big({{H\Psi(x)}\over {\Psi(x)}} \Big),
\end{equation}

\begin{equation}
{\cal S} \equiv Sup \Big({{H\Psi(x)}\over {\Psi(x)}}\Big).
\end{equation}
\vfil\break
Define the configurations:

\begin{equation}
  L_\Psi(x;\lambda_l) \equiv {{H\Psi(x)}\over {\Psi(x)}}  - \lambda_l \geq 0, \iff 
 \lambda_l \leq {\cal I}, 
\end{equation}
and
\begin{equation}
U_\Psi(x;\lambda_u) \equiv \lambda_u - \Big({{H\Psi(x)}\over {\Psi(x)}} \Big ) \geq 0, \iff \lambda_u \geq {\cal S}.
\end{equation}

Although the trial functions must be  positive (i.e. strictly positive) if they
are to be easily used within Barta's procedure, the $\{U_\Psi(x), L_\Psi(x)\}$ functions
can be nonnegative (provided the zeroes correspond to sets of zero measure)
 and still generate strictly positive HH determinants. Accordingly,

\begin{equation}
(H-\lambda_l)\Psi(x) \geq 0, \iff \lambda_l \leq {\cal I},
\end{equation}
and
\begin{equation}
(\lambda_u - H)\Psi(x) \geq 0, \iff \lambda_u \geq {\cal S}.
\end{equation}

Let us focus on the first relation:

\begin{equation}
\Phi_{\lambda_l}(x) = (H-\lambda_l)\Psi(x) \geq 0,\ \lambda_l \leq {\cal I}.
\end{equation}

For any  $\Psi(x) \in {\cal C}$,
 $\Phi_{\lambda_l}(x)$ must be integrable and
positive almost everywhere (i.e. nonnegative). Thus, its power moments must satisfy the standard positivity
relations of the {\it Moment Problem} (Shohat and Tamarkin (1963)), as discussed 
in the Introduction.

We are restricting our analysis to hamiltonians with
rational fraction potentials, since these are the ones most
easily transformable into a moment equation representation. 
For simplicity, the following
discussion  assumes that the potential is of (multidimensional)
polynomial form. The generalization to singular potentials is
straightforward, and briefly discussed below. 

In order to make our analysis more transparent, we
will consider the case of the quartic potential problem:
 $H = -\partial_x^2 + x^4$.
Then $\Phi_{\lambda_l}(x) = (-\partial_x^2 + x^4 - \lambda_l) \Psi(x)$, and we can generate 
the power moments of the LHS, based on those of $\Psi(x)$.

Define the power moments of the trial function by: 
$\mu_p \equiv \int_{-\infty}^{+\infty}dx \ x^p\Psi(x), \ p\geq 0$.
By assumption (i.e. $\Psi > 0$), these must satisfy the Hankel-Hadamard (HH) determinantal 
constraints for the Hamburger moment problem:
$\Delta_{m,n}\Big( \mu \Big ) > 
 0, \ \ for \ m = 0, n\geq 0$.
These constraints are required for all positive (more generally, nonnegative) functions on the real axis.

The power moments of $\Phi_{\lambda_l}(x)$, 

\begin{equation} \nu_p \equiv \int_{-\infty}^{+\infty}dx \ x^p\Phi_{\lambda_l}(x), \ p\geq 0,
\end{equation}
satisfy (i.e. upon substituting the $\Phi/\Psi$ relation and performing the
necessary integration by parts)
\begin{equation}
\nu_p = -p(p-1)\mu_{p-2} + \mu_{p+4} - \lambda_l \mu_p, \ p \geq 0.
\end{equation}
If $\lambda_l \leq {\cal I}$, then the $\nu$ moments generate the Hankel
matrix that must satisfy the (HH) positivity 
constraints:

\begin{equation}
\Delta_{0,N} \Big( \nu \big(\lambda_l \big) \Big) = Det \pmatrix{\cdots \cr
-(n_1 + n_2)(n_1+n_2-1)\mu_{n_1+n_2-2} + \mu_{n_1+n_2+4} - \lambda_l\ \mu_{n_1+n_2} \cr \cdots \cr} > 0,
\end{equation}
for all $0 \leq n_1,n_2 \leq N < \infty$. 

The form of the finite dimensional Hankel matrix in Eq.(30) is symbolized by
{\bf H} -$\lambda_l${\bf U}, 
with ${\bf U}$ the positive definite Hankel matrix for $\Psi$'s moments.

We note that  the $\{\nu_0,\ldots,\nu_{2N}\}$ moments  used to define the Hankel moment matrix
for $\Phi_{\lambda_l}$, depend on the $\{\mu_0,\ldots, \mu_{2N+4}\}$ moments of $\Psi$.
Thus, we are working within the moment space ${\cal U}_Q$ where $Q = 2N+4$. Notational
consistency would suggest that in the following discussion we make reference to
$\lambda_{min;Q_N}(\mu)$ where, $Q_N \equiv 2N+4$. To streamline the discussion,
we will simply use the notation $\lambda_{min;N}$.

Define by $\lambda_{min;N}(\mu)$ the smallest zero satisfying
(the $\mu$ dependence is not explicitly given, for greater clarity)

\begin{equation}
\Delta_{0,N}\Big(\nu\big(\lambda_{min;N}\big)\Big) = 0,
\end{equation}
or
\begin{equation}
Det \Big({\bf H}(\mu)
 - \lambda_{min,N} {\bf U}(\mu)\Big) = 0.
\end{equation}
Accordingly, $(-\infty, \lambda_{min;N}(\mu)) \supset (-\infty,{\cal I})$. This is because, so long as $\lambda_l \leq {\cal I}$, Eq.(30) must hold. Hence, any
root in the $\lambda$-variable domain must be larger than ${\cal I}$.

We now show that the sequence  $\{\lambda_{min,N}(\mu)| N \geq 0\}$, must be 
nonincreasing, or $\lambda_{min,N}(\mu) \geq \lambda_{min,N+1}(\mu)$. The easiest
way is to exploit the positive definiteness of ${\bf U}$, which leads to the
Cholesky decomposition ${\bf U} = {\bf R}^t{\bf R}$, where ${\bf R}$ is the
unique, upper triangular matrix, with positive diagonal entries. Its inverse
is also of upper triangular form. Accordingly, $\lambda_{min;N}$ is also
the smallest zero for the equation
\begin{equation}
Det \pmatrix{{\bf R}^{-t}{\bf H}{\bf R}^{-1} - \lambda_{min,N} {\bf 1}\cr} = 0.
\end{equation}
Of course, this is also the smallest eigenvalue of  the indicated real, symmetric
matrix, which, in turn, define a nonincreasing sequence; thus proving the previous claim.

With regards to the $\lambda_u$'s, an analogous result follows.
Thus, if $\lambda_u \geq {\cal S}$, then
\begin{equation}
\Delta_{0,N}\Big(-\nu\big(\lambda_u \big)\Big) = Det \pmatrix{\cdots \cr
    \lambda_u\ \mu_{n_1+n_2}                   +(n_1 + n_2)(n_1+n_2-1)\mu_{n_1+n_2-2} - \mu_{n_1+n_2+4} 
 \cr \cdots \cr} > 0,
\end{equation}
for all $0 \leq n_1,n_2 \leq N < \infty$.

We now define  ${\lambda_{max;N}(\mu)}$ as the largest root satisfying

\begin{equation}
\Delta_{0,N}(-\nu(\lambda_{max;N})) = 0,
\end{equation}
or
\begin{equation}
Det\big( \lambda_{max,N}{\bf U}(\mu) - {\bf H}(\mu)\Big) = 0.
\end{equation}
(Note that $\lambda_{max;N}$
is also the largest root of the Generalized Eigenvalue Problem, $Det\Big( {\bf H}(\mu) -\lambda_{max,N}{\bf U}(\mu)
\Big) = 0$.)
It then follows that $(\lambda_{max;N}, + \infty) \supset ({\cal S}, +\infty)$,
and they form a  non-decreasing sequence: $\lambda_{max;N}(\mu) \leq \lambda_{max;N+1} (\mu)$. This follows from the observation that $-\lambda_{max;N}$ is
the smallest root of
\vfil\break $Det\Big( -{\bf H}(\mu) -(-\lambda_{max,N}){\bf U}(\mu) \Big) = 0$, and  through the Cholesky decomposition of the positive ${\bf U}$
matrix, the $-\lambda_{max;N}$ form a non-increasing sequence of smallest 
eigenvalues for the finite and symmetric matrix:
 ${\bf R}^{-t}{\bf {(- H)}}{\bf R}^{-1}$.

\subsection {\bf Theorem \# 3}

The first part of Theorem \#3 ( Eq.(14) and Eq.(15)) follow from the
previous results. The latter part of Theorem \#3 (Eq.(16))
results from the fact that the only $\lambda_l$ values
satisfying all of the HH positivity inequalities are those
obeying $\lambda_l \leq {\cal I}$, similarly for $\lambda_u$:
$\lim_{n\rightarrow \infty} \lambda_{min;n}(\mu) = {\cal I}$, and 
$\lim_{n\rightarrow \infty} \lambda_{max;n}(\mu) = {\cal S}$.

\subsubsection {Alternative Derivation}

We include some additional remarks that provide 
a different perspective on all the above. We limit the
discussion, for brevity, to the $\lambda_{min;n}$ case.

The HH positivity theorems require that all of the HH determinants, for the
$\nu$-moments,
be positive, as functions of $\lambda_l$.
 To simply investigate the positivity properties
of one of these determinants is insufficient.  Thus,  we
can either work with 
the quadratic form inequality in Eq.(5), adapted to the
$\nu_p(\mu,\lambda_l)$ moments (Eq.(29)), for the $(N+1) \times  (N+1)$
dimensional Hankel matrix, ${\cal H}_N(\nu)$:

\begin{equation}
\langle {\overrightarrow C}| \pmatrix { \nu_0(\mu,\lambda_l), \nu_1(\mu,\lambda_l), \ldots, \nu_N(\mu,\lambda_l) \cr
\nu_1(\mu,\lambda_l),\nu_2(\mu,\lambda_l), \ldots, \nu_{N+1}(\mu,\lambda_l) \cr
\cdots \cr
\nu_{N}(\mu,\lambda_l), \nu_{N+1}(\mu,\lambda_l), \ldots, \nu_{2N}(\mu,\lambda_l) \cr } | {\overrightarrow C} \rangle > 0,
\ \forall {\overrightarrow C} \neq {\overrightarrow 0}, \end{equation}

\smallskip\noindent or we can work with the $N+1$ HH determinants:

\begin{equation}
\Delta_{0,n}\Big( \nu(\mu,\lambda_l) \Big) > 0, \ n = 0,\ldots,N.
\end{equation}

The set of $\lambda_l$ values satisfying Eq.(37) must correspond to a convex 
set (Chvatal (1983)) since it represents an infinite set of linear inequalitites in the $\lambda$-variable. Thus, the
feasibility $\lambda$ set    must be a semi-infinite interval. 
That is, if Eq.(37) is satisfied by
the two values $\lambda_l =\lambda_l^{(\sigma_{1,2})}$, 

\begin{equation}
\langle {\overrightarrow C}| -(n_1+n_2)(n_1+n_2-1) \mu_{n_1+n_2-2}
+\mu_{n_1+n_2+4} - \lambda_l^{(\sigma_1)} \mu_{n_1+n_2}| {\overrightarrow C}
\rangle > 0,
\end{equation}
\begin{equation}
\langle {\overrightarrow C}| -(n_1+n_2)(n_1+n_2-1) \mu_{n_1+n_2-2}
+\mu_{n_1+n_2+4} - \lambda_l^{(\sigma_2)} \mu_{n_1+n_2}| {\overrightarrow C}
\rangle >0,
\end{equation}
 then it must be satisfied by all
$\lambda_l = s\lambda_l^{(\sigma_{1})} + (1-s)\lambda_l^{(\sigma_{2})}$,
for $0 \leq s \leq 1$ (simply multiply each of the above
two inequalities by $s \geq 0$ and $1-s \geq 0$, respectively, and add); thereby establishing the convex nature of the
set of allowed $\lambda_l$ values.

Since $ \lambda_l \in (-\infty,{\cal I})$ satisfies Eq.(37) (From Eq.(25)),
it follows that there exists a ``largest'' semi-infinite interval $(-\infty, \lambda_{l;N})$, with ${\cal I} \leq \lambda_{l;N} $, satisfying all of Eq.(37). 
It is clear that $\lambda_{l;N+1} \leq \lambda_{l;N}$, since from Eq.(37),
the quadratic form inequalities for the ${\cal H}_{N+1}(\nu)$ Hankel matrix
includes all of those corresponding to the ${\cal H}_N(\nu)$ case. The
$\lambda_{l;N}$ value must then be the smallest root of the equation
$Det({\cal H}_N(\nu(\mu,\lambda_l)) = 0$. That is, $\lambda_{l;N} = \lambda_{min;N}$.

\subsubsection{Extension to Rational Fraction (Singular) Potentials}

The preceding, alternative, proof also shows us how to extend our results
to the case of  rational fraction type potentials.
Consider the perturbed quartic potential $V(x) = x^4 + {1\over{x^2+2}}$. Limiting ourselves to the ``infimum''case, for simplicity,
we see that Eq.(27) can be modified by multiplying both sides by the
positive denominator polynomial, $x^2+2$:

\begin{equation}
(x^2+2)\Phi_{\lambda_l}(x) = \Big(-(x^2+2)\partial_x^2 + (x^2+2) x^4 + 1 - \lambda_l(x^2+2)\Big)\Psi(x) > 0, \ \lambda_l \leq {\cal I}.
\end{equation}
Thus, the R.H.S. generates a positive (Hankel) matrix, and one can procede to
define the corresponding $\lambda_{min;N}$, which satisfies all the relations
described above.

Thus, in general, as long as one multiplies Eq.(27), or its multidimensional 
counterpart, by positive ``denominator'' type polynomials, in order to
achieve a Hankel matrix structure, then all of our results apply.

\subsubsection{Additional Remarks  }

Given that the function ${{H\Psi}\over \Psi}$  can have
multiple extrema, having a systematic method of computing
the infimum/supremum (as opposed to searching over all local extrema) 
may make the above results very convenient. That is, 
one can determine  the infimum (${\cal I}(\mu)$) and supremum (${\cal S}(\mu)$)
 by
studying the asymptotic limits of the $\lambda_{min;n}$ and $\lambda_{max;n}$,
combined with sequence acceleration techniques, where possible.

Another important aspect of the previous results is that we can now extend
Barta's result to positive functions which may not be given in closed form, but
whose moments may be known. We provide one example of this in the next section.

\subsection {\bf Theorem \# 2}

We now prove Theorem \#2 by way of the quartic potential problem. 
Whereas in the previous proofs we were working with an infinite
set of numbers, $\{\mu_p| p \geq 0\}$, known to be the moments of a
positive function, we will now be working with a finite set of
moments that satisfy the Moment Equation, as well as the corresponding
positivity theorems.

Let us assume that the $\{\mu_p| 0 \leq p \leq P\}$ moments satisfy the moment equation,

\begin{equation}
-p(p-1)\mu_{p-2} + \mu_{p+4} - E \mu_p = 0, \ 0 \leq p \leq P-4.
\end{equation}
Alternatively,
\begin{equation}
 \mu_{p+4} = E \mu_p + p(p-1)\mu_{p-2}, \ 0 \leq  p \leq P-4.
\end{equation}
Clearly, this recursive relation separates into the even and odd order
moments. There are more efficient ways of dealing with such relations, however,
for our immediate purposes, the above is satisfactory. Also, we
implicitly assume that some normalization has been chosen.

Let $P = 2M$. Again, we assume that
the moments $\{\mu_0,\mu_1,\ldots,\mu_{P=2M}\} \in {\cal U}_{P=2M;EMM} \subset 
{\cal U}_{P=2M}$,
 satisfy the moment 
equation and all the HH determinantal inequality conditions that
can be generated from them, for some $E$ value. Thus 
$\Delta_{0,n}(\mu) > 0$, for $n \leq M$.
From Eq.(30), we see that the ${\bf U}$ matrix
 involves all of the moments up to order $\mu_{2N}$, where $N$
is to be determined.
The ${\bf H}$ matrix involves the highest order moment,
$\mu_{2N+4}$. Thus, we want $2N+4 = 2M$. That is, the highest
dimension Generalized Eigenvalue Problem (GEP) is $N+1 = M-1$.

From Eq.(42), it follows that for the special set of 
moments being considered, we have  ${\bf H} = E {\bf U}$. The
corresponding  GEP problem  becomes

\begin{equation}
Det\Big( {\bf H} - \lambda {\bf U} \Big) = 
Det\Big( E {\bf U} - \lambda {\bf U} \Big) =  (E-\lambda)^{N+1} Det\Big({\bf U}
\Big),
\end{equation}
where $Det\Big({\bf U}\Big) = \Delta_{0,N}(\mu) $,
revealing its $(N+1)$-th order degeneracy. Hence

\begin{equation}
\lambda_{min;N}(\mu_E) = \lambda_{max;N}(\mu_E).
\end{equation}
In summary, the GEP problem becomes extremely degenerate for
those moments satisfying the Moment Equation, as well
as all of the corresponding HH positivity constraints.
The allowable $E$ values are those generated through the
EMM procedure corresponding to moment order $P = 2M$.

\subsection {\bf Theorem \# 4 }

Define the supremum of the smallest GEP eigenvalue by
$\lambda^{sup}_{min;Q} = Sup_{\mu \in {\cal U}_Q}\lambda_{min;Q}(\mu)$;
and the infimum of the largest GEP eigenvalue by
$\lambda^{inf}_{max;Q} = Inf_{\mu \in {\cal U}_Q}\lambda_{min;Q}(\mu)$.
Since the EMM related set of moments satisfy
 ${\cal U}_{Q;EMM} \subset {\cal U}_{Q}$, and on ${\cal U}_{Q;EMM}$
the extremal eigenvalues
are degenerate, it follows
 that
the EMM upper bound, must be a lower bound to $\lambda^{sup}_{min;Q}$.
Likewsize, the EMM lower bound, must be an upper bound
to $\lambda^{inf}_{max;Q}$. This confirms Eq.(18).

In the $Q\rightarrow \infty$ limit, the entire moment
space, for a given set of moments ${\overrightarrow \mu}
= (\mu_0,\ldots, \mu_{j\rightarrow \infty})$, we must
have that $\lambda_{min;\infty}(\mu) < E_{gr} < 
\lambda_{max;\infty}(\mu)$, from Eq.(16). Clearly then
$\lambda^{sup}_{min;\infty} = E_{gr} = \lambda^{inf}_{max;\infty}$;
confirming Eq.(19).

Again, we implicitly assumed that ${\cal U}_Q$ satisfies some physically
motivated normalization prescription.

\vfil\break
\section {Some Numerical Results}

We will use the quartic potential to illustrate, numerically,
most of the previous results. The ground state energy is
$E_{gr} =  1.060362090484 $.

\subsection {Eqs.(14-16), using $\Psi(x) = {\cal N} e^{-x^2 }$}

This is a trivial example. One has ${{H\Psi(x)}\over {\Psi(x)}} = 
2-4x^2+x^4$. The infimum is ${\it Inf}\Big(2-4x^2+x^4\Big) = -2$.
The even order, gaussian function power moments,
satisfy the recursion relation  $\mu_{p+2} = \Big({{1+p}\over 2}\Big) \mu_p$,
$p \geq 0$. Normalizing according to  $\mu_0 \equiv 1$, determines the
normalization factor ${\cal N}$.

Since the Gaussian function, $\Psi_g = e^{-x^2}$ also satisfies 
$-\partial_x^2\Psi_g +4x^2\Psi_g = 2 \Psi_g$, an alternate recursion relation for the moments is 
$-p(p-1)\mu_{p-2} + 4 \mu_{p+2} = 2 \mu_p$. Using this, one can transform
the matrix structure  in Eq.(30) (i.e.  $-p(p-1)\mu_{p-2} + \mu_{p+4} - \lambda \mu_{p}$) into
 $ -4\mu_{p+2} + \mu_{p+4} - (\lambda -2) \mu_p$. The Generalized Eigenvalue Problem results, for this problem, are given in Table I. Note that
the convergence is slow, but consistent with the various Theorems. We also 
note the curious repetitive, non-repetitive, structure manifested by the
eigenvalues. No sequence acceleration analysis has been
attempted.

\subsection {Eqs.(14-16), using $\Psi(x) = |\Phi(x)|^2$, where
$\Phi(x)$ satisfies the PT-invariant Schrodinger equation
$-\partial_x^2\Phi -(ix)^3\Phi = {\cal E} \Phi$}

We now investigate the utility of the previous formalism when the
positive trial function is not known, in closed form; although the
moments are (numerically) known. For this excersize, we could  take
$\Psi(x)$ to be the (positive) ground state of any ({\underline s}olvable)
Schrodinger potential problem, $-\partial_x^2\Psi_s + V_s(x) \Psi_s(x) = E_s \Psi_s(x)$. In such cases, one would determine the {\it Inf}/{\it Sup} of (i.e. $H_4 \equiv -\partial_x^2 + x^4$),
 ${{H_4\Psi_{s}}\over {\Psi_{s}}} = -{{\partial_x^2\Psi_{s}}\over {\Psi_{s}}} + x^4 = E_s - V_s(x) + x^4$. This would then be a trivial analysis.

Instead, we pursue a different class of problems whose differential structure
does not lead to an easily calculable set of Barta bounds. Such is provided
by the class of non-hermitian systems that have received much attention in
the context of PT-symmetry breaking systems (Bender and Boettcher (1998)). The simplest example of this is
the well known $V(x) = -(ix)^3$ system, which we write as $-\partial_x^2\Phi(x) -(ix)^3\Phi(x) = {\cal E}\Phi(x)$.  This system admits only real eigenenergies;
however, its eigenstates are all complex functions. Nevertheless, the probability density, $\Psi(x) = |\Phi(x)|^2 \equiv S(x) >0$, satisfies a linear, fourth 
order differential equation (Handy (2001)):

\begin{equation}
\partial_x\Big( -{1\over {x^3}} \partial_x^3 S(x) - 4{{\cal E}\over{x^3}}\partial_xS(x) \Big) + 4x^3 S(x) = 0.
\end{equation}

Although all of the bound states of this system are positive, we shall work with the one corresponding to the smallest ${\cal E} = 1.1562670719881133$. The
Hamburger moments of the even function, $S(x)$, satisfy a simple recursion relation (Handy (2001)):

\begin{equation}
4\mu_{p+7} = (p+4)p(p-1)(p-2)\mu_{p-3}
 +   4{\cal E}p(p + 4)  \mu_{p-1},
\end{equation}
for $p \geq 0$.
The GEP-moment analysis, given in Table II tells us that
${\it Inf}\Big({{H_4S(x)}\over {S(x)}}\Big) < -1.57$.
 In order to verify
this, we can implement a Runge-Kutta analysis on $S(x)$, in order
to calculate $-{{S''}\over S}$. If one is not too careful (i.e. implementation 
of a naive second order finite differencing), a significantly
wrong answer is obtained (i.e. Barta's infimum is $O(.3)$). Instead, by using
the relation
 $\Big({{H_4S(x)}\over {S(x)}}\Big) = 
- {{\Phi\partial_x^2\Phi^* + \Phi^*\partial_x^2\Phi + 2 \partial_x\Phi^* \partial_x\Phi}\over {\Phi^* \Phi}} + x^4 = 2{\cal E} - 2|{{\partial_x\Phi}\over \Phi}|^2 + x^4$, the resulting expression lends itself to a more accurate Runge-Kutta
verification, yielding the (approximate) Barta {\it infimum} as $-1.782$. This is very
consistent with the GEP generated results in Table II.

\subsection{Generating Converging Bounds for Quartic Potential (Theorem \# 4):\\ The need for a rough upper bound to the energy}

In this last example, we will not work with a fixed set of moments for a 
positive trial configuration. Instead, we will implement an optimization
procedure for determining $Sup\Big(\lambda_{min;Q}(\mu)\Big)$ and
$Inf\Big(\lambda_{max;Q}(\mu)\Big)$, for $\mu \in {\cal U}_Q$.
Contrary to the Gradient analysis in Mouchet's (2005) work, we can determine these
quantities by combining the
 EMM, linear programming based,  ``Cutting-Algorithm" (Handy et al (1988))
 with a bisection method in the
$\lambda$-variable space. 

\subsubsection {Defining ${\cal U}_Q$}

The ${\cal U}_Q$ space is defined as the set of Hamburger 
moments,$\{(\mu_0,\ldots,\mu_Q)\}$, satisfying

\begin{equation}
\langle {\overrightarrow C_1}| 
 \mu_{n_1+n_2}| {\overrightarrow C_1}
\rangle > 0,\ \forall {\overrightarrow C_1} \neq 0,\  0 \leq n_1+n_2 \leq Q.
\end{equation}
In addition, for the case of $Sup\Big(\lambda_{min;Q}(\mu)\Big)$,
 we are interested in the set of $\lambda_l$'s satisfying (i.e. Eq.(30))

\begin{equation}
\langle {\overrightarrow C_2}|-(n_1+n_2)(n_1+n_2-1) \mu_{n_1+n_2-2}
+\mu_{n_1+n_2+4} - \lambda_l \mu_{n_1+n_2}| {\overrightarrow C_2}
\rangle > 0, \forall {\overrightarrow C_2} \neq 0, 
\end{equation}
\smallskip\noindent $ 0 \leq n_1+n_2+4 \leq Q$.

Whereas, for the $Inf\Big(\lambda_{max;Q}(\mu)\Big)$, the latter
set of inequalities are replaced by (i.e. Eq.(34))

\begin{equation}
\langle {\overrightarrow C_2}| 
 {\lambda_u} \mu_{n_1+n_2}
+(n_1+n_2)(n_1+n_2-1) \mu_{n_1+n_2-2} -\mu_{n_1+n_2+4}|
 {\overrightarrow C_2} \rangle
 > 0, \forall {\overrightarrow C_2} \neq 0,
\end{equation}
\smallskip\noindent $ 0 \leq n_1+n_2+4 \leq Q$.

\subsubsection {Normalization Prescription: Bounding ${\cal U}_Q$}

One must also impose some normalization condition.
A choice that leads to a bounded ${\cal U}_Q$ set is:

\begin{equation}
\mu_0 + \mu_{Q(even)} = 1.
\end{equation}
To study the consequences of this, note that the physical moments for the 
ground state wavefunction (i.e. assume $\Psi = \Psi_{gr}$)
  satisfy
\begin{equation}
\mu_p  = \int_{-1}^{+1}dx \ x^p \Psi(x) + \int_{x\notin [-1,1] } dx \ x^p  \Psi(x).
\end{equation}
The $p =\ {\rm {even}}$ moments must be positive and satisfy
\begin{equation}
0 < \mu_{p = even} =  \int_{-1}^{+1}dx\ \ x^p \Psi(x) + \int_{x\notin [-1,1] } dx \ x^p\Psi(x) < \int_{-1}^{+1}dx\ \  \Psi(x)  + \int_{x\notin [-1,1] } d x \ x^Q\Psi(x),
\end{equation}
or
\begin{equation}
0 < \mu_{p = even}  < \mu_0 + \mu_Q = 1.
\end{equation}
For the odd order moments, a similar set of relations ensues for $|\mu_{p=odd}| \leq \int_{-1}^{+1}dx\ \ |x^p| \Psi(x) + \int_{x\notin [-1,1] } dx\ |x^p|\Psi(x)< \mu_0 + \mu_Q = 1$. Thus, we have

\begin{equation}
-1 \leq \mu_{p = odd} \leq +1.
\end{equation}

\subsubsection {Linear Programming - Bisection Algorithm for Determining $\lambda_{min}^{Sup}$ and $\lambda_{min}^{Sup}$}

The following algorithm implicitly  makes use of the quasi-convex nature of $\lambda_{min;Q}(\mu)$ and the quasi-concave 
structure of $\lambda_{max;Q}(\mu)$ for $\mu \in {\cal U}_Q$.

We outline the basic structure of our computational algorithm.
Assume that for a 
trial positive solution, $(\mu^*_0,\ldots,\mu^*_Q)$, we have determined its
corresponding extremal eigenvalue, $\lambda_{min;Q}(\mu_*)$.
Within the interval $[\lambda_{min;Q}(\mu_*), \infty)$, we 
pick an arbitrary point, $\lambda_a$, and use the EMM ``Cutting-
Algorithm" to determine if there exists a point in ${\cal U}_Q$, satisfying the
normalization conditions, as well as
 Eq.(49), for $\lambda_l = \lambda_a$. There are two possibilities:

(A) If there is such a point, then we repeat
the entire procedure, but within the interval $[\lambda_a, +\infty)$.

(B) If there is no such point, then the entire procedure is repeated
within the interval $[\lambda_{min;Q}(\mu_*), \lambda_a]$.

The objective is to eventually generate a reducing sequence of intervals,
$[ \lambda_{a_1},\lambda_{a_2}]\supset \cdots \supset [ \lambda_{a_i},\lambda_{a_{i+1}}]$, until an acceptably small interval is attained. The 
endpoints will tightly bound $\lambda_{min;Q}^{Sup}$.

For the $\lambda_{max;Q}^{Inf}$, a similar procedure is required. Thus,
one would select a point within the interval $(-\infty, \lambda_{max;Q}(\mu_*))$
. Upon picking an arbitrary point within this interval, $\lambda_a$, one would then determine the existence, or non-existence of a $\mu$-point lying within ${\cal U}_Q$, and satisfying the normalization conditions. Such a point must 
also satisfy Eq.(50), for $\lambda_u = \lambda_a$. If there is such a $\mu$-point, then the entire procedure is repeated for the interval $(-\infty, \lambda_a)$. If there is no such point, then the updated interval is $(\lambda_a,\lambda_{max;Q}(\mu_*))$.

\subsubsection { The Need for a Rough Upper Bound}

All of the above is contingent on making sure that the boundary of ${\cal U}_Q$ include no points at which the ${\bf U}(\mu)$ matrix
has zero eigenvalues. This was emphasized previously. The adopted choice of normalization, if not supplemented with additional 
linear constraints on the moments, include such singular points. Specifically, because the ${\bf U}$ matrix only includes moments
up to order $\mu_{2N}$, while the ${\bf H}$ matrix includes the additional moments $\{\mu_{2N+1}, \ldots, \mu_{2N+4}\}$,  
one possible boundary point could be all of the first $2N+4$ moments set to zero (i.e. $\mu_{0\leq n \leq 2N+3} = 0$) and the
last moment set to unity, $\mu_{2N+4} = 1$. To avoid these, and other such possibilities, any rough upper bound for $E_{gr}$
will help in restricting ${\cal U}_Q$ to avoid such boundaries. 

Let $E_{pub} >> E_{gr}$ denote a {\it poor  upper bound} to the ground state energy. The true moment equation for the quartic problem is
$ E_{gr} \mu_p = -p(p-1)\mu_{p-2} + \mu_{p+4}$. Taking $ p = even $, we have

\begin{equation}
E_{pub}\ \mu_p + p (p-1) \mu_{p-2} > \mu_{p+4}, \  p = {\rm even}.
\end{equation}
Thus, these additional inequality relations  will lead to proper ${\cal U}_Q$ sets. In Table III, we take $E_{pub} = 2$. 

The results in Table III confirm the all of the above theoretical results. Note that the EMM bounds will be, generally, tighter
than those derived from a ``moment problem extension of Barta's theorem''. The calculations were done using the Stieltjes form
for the moments (i.e. all odd  order Hamburger moments were set to zero, abinitio, $\mu_{odd} = 0$). The results in Table III
confirm that knowledge of a rough upper bound for the ground state energy lead to converging bounds for the ground state energy.
This result is similar to that developed in a Euclidean time reformulation of the EMM philosophy, as applied to positive matrices
(Handy and Ndow (1992)).

\begin{table}
\caption {Quartic potential results using ${\cal N}e^{-x^2}$ trial function;
Barta's lower bound is -2 (Note that ${\it Dim}  \equiv N+1$ and Max. Moment Order, $Q$,  satisfy $Q = 4+2N$)
}
\begin{center}
\begin{tabular}{cccccl}
 \multicolumn{1}{c}{$ Dim (Q)$}
& \multicolumn{1}{c}{$\lambda_{min:Q}$} & \multicolumn{1}{c}{$Dim (Q)$}
&\multicolumn{1}{c}{$\lambda_{min:Q}$}&\multicolumn{1}{c}{$ Dim (Q)$}
& \multicolumn{1}{c}{$\lambda_{min:Q}$} \\ \hline
1 (  4) &    .75000  &  34 ( 70) &  -1.74204  & 67 (136) &  -1.86361 \\
  2 (  6) &   -.25000  &  35 ( 72) &  -1.75151  & 68 (138) &  -1.86687 \\
  3 (  8) &   -.25000  &  36 ( 74) &  -1.75151  & 69 (140) &  -1.86687 \\
  4 ( 10*) &  -.45810  &  37 ( 76*)&  -1.75380  & 70 (142) &  -1.86840 \\
  5 ( 12) &   -.82522  &  38 ( 78) &  -1.75447  & 71 (144) &  -1.86840 \\
  6 ( 14) &   -.82522  &  39 ( 80) &  -1.75447  & 72 (146) &  -1.8688004 \\
  7 ( 16) &  -1.06261  &  40 ( 82) &  -1.77664  & 73 (148) &  -1.8688004 \\
  8 ( 18) &  -1.06261  &  41 ( 84) &  -1.77664  & 74 (150*) &  -1.8688013 \\
  9 ( 20*)&  -1.06705  &  42 ( 86) &  -1.79200  & 75 (152) &  -1.87513 \\
 10 ( 22) &  -1.28893  &  43 ( 88) &  -1.79200  & 76 (154) &  -1.87513 \\
 11 ( 24) &  -1.28893  &  44 ( 90) &  -1.80067  & 77 (156) &  -1.88065 \\
 12 ( 26) &  -1.37898  &  45 ( 92) &  -1.80067  & 78 (158) &  -1.88065 \\
 13 ( 28) &  -1.37898  &  46 ( 94) &  -1.80422  & 79 (160) &  -1.88475 \\
 14 ( 30*)&  -1.38656  &  47 ( 96) &  -1.80422  & 80 (162) &  -1.88475 \\
 15 ( 32) &  -1.46259  &  48 ( 98*)&  -1.80483  & 81 (164) &  -1.88746 \\
 16 ( 34) &  -1.46259  &  49 (100) &  -1.80920  & 82 (166) &  -1.88746 \\
 17 ( 36) &  -1.54181  &  50 (102) &  -1.80920  & 83 (168) &  -1.88894 \\
 18 ( 38) &  -1.54181  &  51 (104) &  -1.82258  & 84 (170) &  -1.88894 \\
 19 ( 40) &  -1.56636  &  52 (106) &  -1.82258  & 85 (172) &  -1.88953 \\
 20 ( 42) &  -1.56636  &  53 (108) &  -1.83219  & 86 (174) &  -1.88953 \\
 21 ( 44*)&  -1.56786  &  54 (110) &  -1.83219  & 87 (176*)&  -1.88962 \\
 22 ( 46) &  -1.61360  &  55 (112) &  -1.83802  & 88 (178) &  -1.89137 \\
 23 ( 48) &  -1.61360  &  56 (114) &  -1.83802  & 89 (180) &  -1.89137 \\
 24 ( 50) &  -1.65766  &  57 (116) &  -1.84077  & 90 (182) &  -1.89588 \\
 25 ( 52) &  -1.65766  &  58 (118) &  -1.84077  & 91 (184) &  -1.89588 \\
 26 ( 54) &  -1.67629  &  59 (120) &  -1.841515  & 92 (186) &  -1.89958 \\
 27 ( 56) &  -1.67629  &  60 (122) &  -1.841515  & 93 (188) &  -1.89958 \\
 28 ( 58*)&  -1.68012  &  61 (124*)&  -1.841519  & 94 (190) &  -1.90236 \\
 29 ( 60) &  -1.68637  &  62 (126) &  -1.85045  & 95 (192) &  -1.90236 \\
 30 ( 62) &  -1.68637  &  63 (128) &  -1.85045  & 96 (194) &  -1.90422 \\
 31 ( 64) &  -1.72107  &  64 (130) &  -1.85818  & 97 (196) &  -1.90422 \\
 32 ( 66) &  -1.72107  &  65 (132) &  -1.85818  & 98 (198) &  -1.90529 \\
 33 ( 68) &  -1.74204  &  66 (134) &  -1.86361  & 99 (200) &  -1.90529 \\
  & &   &       &  100 (202) &  -1.90576\\
  & &   &       &  101 (204) &  -1.90576\\
\end{tabular}
\end{center}
\noindent{* Except for these entries, all others appear in pairs, to 20 significant figures}
\end{table}

\vfil\break
\begin{table}
\caption {Quartic potential results using as trial configuration the $|\Phi(x)|^2$ solution corresponding to  PT-invariant, non-hermitian, system $-\partial_x^2\Phi(x) -(ix)^3\Phi(x) = {\cal E} \Phi(x)$, for ${\cal E} =  1.1562670719881133$.
Barta's lower bound is {\it approximately} (-1.782) , based upon Runge-Kutta integration.
}
\begin{center}
\begin{tabular}{cccccl}
 \multicolumn{1}{c}{$ Q$}
& \multicolumn{1}{c}{$\lambda_{min:Q}$} & \multicolumn{1}{c}{$Q$}
&\multicolumn{1}{c}{$\lambda_{min:Q}$}
\\ \hline
4 & 0.7651830316 & 30 &-1.412946343 \\
 6 & -0.3701497316 & 32& -1.412946343\\
 8 & -0.5797495842 & 34& -1.466405630\\
 10 & -0.5797495842 & 36& -1.466405630\\
 12 & -0.9175699949 & 38& -1.471430659\\
 14 & -1.025936484 & 40&  -1.505086215\\
 16 & -1.025936484 & 42& -1.505086215\\
 18 & -1.202683806 & 44& -1.535124305\\
 20 & -1.202683806 & 46& -1.535124305\\
 22 & -1.202840090 & 48& -1.530286871\\
 24 & -1.349131584 & 50& -1.556428376\\
 26 & -1.349131584 & 52& -1.556428376\\
 28 & -1.369576335 & 54& -1.579966618\\
&           & 56 & -1.5799666184\\
&           & 58 & -1.5799666184\\
&           & 60 & -1.5882508326\\
\end{tabular}
\end{center}
\noindent{}
\end{table}

\vfil\break
\begin{table}
\caption {Results of the Quasi-Convexity/Concavity Analysis (i.e. ``Moment Problem Reformulation of Barta's Theorem'') Applied
to the Quartic Potential Problem: $-\Psi''(x) + x^4\Psi(x) = E \Psi(x)$.} 
\begin{center}
\begin{tabular}{cccccl}
 \multicolumn{1}{c}{ Moment Order $P^*$}&
 \multicolumn{1}{c}{Theorem \# 4 Bounds} & 
\multicolumn{1}{c}{EMM Bounds} \\ \hline
 6 & $.934 < E_{gr} < 1.170$   &     $.934 < E_{gr} < 1.150$  \\
 7 & $1.021 < E_{gr} <   1.168$ &    $1.028 < E_{gr} < 1.153$ \\
 8 & $1.027 < E_{gr} <    1.080$ &   $1.028 < E_{gr} < 1.067$ \\
 9 & $1.050  < E_{gr} <   1.068$ &   $1.059 < E_{gr} < 1.067$\\
 10 & $1.055  < E_{gr} <  1.063$ &   $1.059 < E_{gr} < 1.062$\\
 11 & $1.055  < E_{gr} <  1.062$ &   $1.059 < E_{gr} < 1.061$\\
 12 & $1.0602 < E_{gr} <  1.0613$&   $1.0602 < E_{gr} < 1.0610$\\
\end{tabular}
\end{center}
\noindent{$P^*: \ \{\mu_{2\rho}| 0 \leq \rho \leq P\}$}
\end{table}

\vfil\break 
\section {Conclusion}

We have outlined a theoretical procedure for transforming Barta's configuration space theorem
into a moment problem equivalent. The advantages of the latter are that it leads to a (quasi)-convexity/concavity
reformulation that avoids multi-extrema difficulties associated with the configuration space formulation.
In addition, by so doing, we solve the problem of defining a procedure for improving Barta's bounds,
once an initial trial configuration is used. This was an outstanding, theoretical problem, within the
configuration space formulation. We show that the Eigenvalue Moment Method (EMM), is an integral
part of this procedure, and allows us to prove Theorem \# 4. In turn, the results presented here
prove that the EMM feasibility energy values correspond to a continuous set (an interval) since it
is bounded by the supremum and infimum of the extremal eigenvalues associated with the underlying 
Generalized Eigenvalue Problem.

\vfil\break
\section {Acknowledgments}
This work benefitted from partial, visitation, support extended to the author through NSF award NBTC-URG 47956-7824, involving
 Cornell University's Nanobiotechnology center and Clark Atlanta University (CAU). The efforts, in this regard, of Dr. Ishrat Khan are gratefully
acknowledged. This work benefitted from discussions with
Dr. Lois Pollack's group focusing on protein folding studies.  Additional insights from discussions with Dr. C. J. Tymczak, Mr. Harold Brooks, 
Professor Daniel Bessis, and Mr. Joshi Siddharth,  as well as the computing resouces of CAU's Center for Theoretical Studies of Physical Systems, are gratefully
acknowledged. 

\vfil\break
\section {References}

\noindent Baker  G A Jr.  {\it Essentials of Pade Approximants in Theoretical Physics} (New York: Academic Press 1975)

\noindent Barnsley M F 1978 J. Phys. A: Math. Gen. {\bf 11} 55

\noindent Barnsley M F {\it Fractals Everywhere} (New York: Academic Press 1988)

\noindent Barta J 1937 C. R. Acad. Sci. Paris {\bf 204} 472

\noindent Bender C M and Boettcher S 1998 Phys. Rev. Lett. {\bf 80} 5243

\noindent Bender C M and Orszag S A,   {\it Advanced Mathematical
Methods for
Scientists and Engineers} (New York: McGraw Hill 1978).

\noindent Boyd S and Vandenberghe L {\it Convex Optimization} (New York: Cambridge University Press 2004)

\noindent Chvatal V 1983 {\it Linear Programming} (Freeman, New York)

\noindent Daubechies I 1988 Comm. Pure \& Appl. Math. {\bf 41} 909

\noindent Devinatz A 1957 Duke Math J. {\bf 24} 481

\noindent Grossmann  A and Morlet J 1984 SIAM J. Math. Anal. {\bf 15} 723 

\noindent Handy C R 1981 Phys. Rev. D {\bf 24} 378

\noindent Handy C R 1984 CAU preprint (unpublished)

\noindent Handy C R 2001 J. Phys. A: Math. Gen. {\bf 34} 5065

\noindent Handy C R and  Bessis D 1985 Phys. Rev. Lett. {\bf 55}, 931

\noindent Handy C R, Bessis D, Sigismondi G, and Morley T D 1988b Phys. Rev. Lett
{\bf 60}, 253

\noindent Handy C R, Giraud B G, and Bessis D 1991 Phys. Rev. A {\bf 44} 1505

\noindent Handy C R, Maweu J, and Atterberry L 1996  J. Math. Phys. {\bf 37} 1182

\noindent Handy C R,  Msezane A Z, and Yan Z 2002 J. Phys. A : Math. Gen. {\bf 35} 6359

\noindent Handy C R and Murenzi R 1997 J. Phys. A: Math. Gen {\bf 30} 4709

\noindent Handy C R and Murenzi R 1998 J. Phys. A: Math. Gen {\bf 31} 9897

\noindent Handy C R and Murenzi R 1999 J. Phys. A: Math. Gen {\bf 32} 8111

\noindent Handy C R and Ndow G L 1992 J. Phys. A: Math. Gen. {\bf 25} 2669 

\noindent Kirkpatrick S, Gelatt C D, and Vecchi M P 1983 Science {\bf 220} 671

\noindent Le Guilou J C and Zinn-Justin J 1983  Ann. Phys. (N.Y.) 147, 57 

\noindent Mouchet A 2005 J. Phys. A: Math. Gen. {\bf 38} 1039

\noindent Shohat J A and  Tamarkin J D, {\it The Problem of Moments}
(American Mathematical Society, Providence, RI, 1963).

\noindent Siddharth J 2005 (private communication)

\noindent Thirumalai D and Hyeon C 2005 Biochemistry  {\bf 44} 4957

\noindent Watkins D S {\it Fundamentals of Matrix Computations} (New York: John Wiley \& Sons, Inc. 2002)
\end{document}